\documentclass[twoside,twocolumn,9pt,colorlinks=true]{article}
\usepackage{extsizes}
\usepackage[super,sort&compress,comma]{natbib} 
\usepackage[version=3]{mhchem}
\usepackage[left=1.5cm, right=1.5cm, top=1.785cm, bottom=2.0cm]{geometry}
\usepackage{balance}
\usepackage{mathptmx}
\usepackage{sectsty}
\usepackage{graphicx} 
\usepackage{lastpage}
\usepackage[format=plain,justification=justified,singlelinecheck=false,font={stretch=1.125,small,sf},labelfont=bf,labelsep=space]{caption}
\usepackage{float}
\usepackage{fancyhdr}
\usepackage{fnpos}
\usepackage[english]{babel}
\usepackage{array}
\usepackage{droidsans}
\usepackage{charter}

\usepackage[TS1,T1]{fontenc}
\usepackage[usenames,dvipsnames]{xcolor}
\usepackage{setspace}
\usepackage[compact]{titlesec}
\usepackage{doi}
\usepackage{hyperref}
\usepackage{tabularx}
\usepackage{siunitx}
\usepackage{amsmath,amssymb}
\usepackage{microtype}
\usepackage{soul}
\usepackage{textcomp}
\usepackage{epstopdf}
\newcommand{\ftextnumero}{{\fontfamily{txr}\selectfont \textnumero}}
\definecolor{cream}{RGB}{222,217,201}
\DeclareSIUnit\Molar{\textsc{m}}

\DeclareMathOperator{\Si}{Si}
\newcommand{\V}{\langle v \rangle}
\newcommand{\D}{\langle D \rangle}
\begin{document}

\pagestyle{fancy}
\thispagestyle{plain}
\fancypagestyle{plain}{
\renewcommand{\headrulewidth}{0pt}
}

\makeFNbottom
\makeatletter
\renewcommand\LARGE{\@setfontsize\LARGE{15pt}{17}}
\renewcommand\Large{\@setfontsize\Large{12pt}{14}}
\renewcommand\large{\@setfontsize\large{10pt}{12}}
\renewcommand\footnotesize{\@setfontsize\footnotesize{7pt}{10}}
\makeatother

\renewcommand{\thefootnote}{\fnsymbol{footnote}}
\renewcommand\footnoterule{\vspace*{1pt}%
\color{cream}\hrule width 3.5in height 0.4pt \color{black}\vspace*{5pt}} 
\setcounter{secnumdepth}{5}

\makeatletter 
\renewcommand\@biblabel[1]{#1}            
\renewcommand\@makefntext[1]%
{\noindent\makebox[0pt][r]{\@thefnmark\,}#1}
\makeatother 
\renewcommand{\figurename}{\small{Fig.}~}
\sectionfont{\sffamily\Large}
\subsectionfont{\normalsize}
\subsubsectionfont{\bf}
\setstretch{1.125} 
\setlength{\skip\footins}{0.8cm}
\setlength{\footnotesep}{0.25cm}
\setlength{\jot}{10pt}
\titlespacing*{\section}{0pt}{4pt}{4pt}
\titlespacing*{\subsection}{0pt}{15pt}{1pt}

\fancyfoot{}
\fancyfoot[LO,RE]{\vspace{-7.1pt}}
\fancyfoot[CO]{\vspace{-7.1pt}}
\fancyfoot[CE]{\vspace{-7.2pt}}
\fancyfoot[RO]{\footnotesize{\sffamily{1--\pageref{LastPage} ~\textbar  \hspace{2pt}\thepage}}}
\fancyfoot[LE]{\footnotesize{\sffamily{\thepage~\textbar ~1--\pageref{LastPage}}}}
\fancyhead{}
\renewcommand{\headrulewidth}{0pt} 
\renewcommand{\footrulewidth}{0pt}
\setlength{\arrayrulewidth}{1pt}
\setlength{\columnsep}{6.5mm}
\setlength\bibsep{1pt}

\makeatletter 
\newlength{\figrulesep} 
\setlength{\figrulesep}{0.5\textfloatsep} 

\newcommand{\topfigrule}{\vspace*{-1pt}%
\noindent{\color{cream}\rule[-\figrulesep]{\columnwidth}{1.5pt}} }

\newcommand{\botfigrule}{\vspace*{-2pt}%
\noindent{\color{cream}\rule[\figrulesep]{\columnwidth}{1.5pt}} }

\newcommand{\dblfigrule}{\vspace*{-1pt}%
\noindent{\color{cream}\rule[-\figrulesep]{\textwidth}{1.5pt}} }

\makeatother

\twocolumn[
\begin{@twocolumnfalse}
\vspace{1em}
\sffamily
\begin{tabular}{m{4.5cm} p{13.5cm} }

~ & \noindent\LARGE{\textbf{Particle sizing for flowing colloidal suspensions using \mbox{flow-differential dynamic microscopy$^{\dag}$}}} \\
\vspace{0.3cm} & \vspace{0.3cm} \\

&\noindent\large{James A. Richards,$^{\ast}$ Vincent A. Martinez,$^{\ast}$ and Jochen Arlt$^{\ast}$} \\ \vspace{1.5cm}

~ & \noindent\normalsize{Particle size is a key variable in understanding the behaviour of the particulate products that underpin much of our modern lives. Typically obtained from suspensions at rest, measuring the particle size under flowing conditions would enable advances for in-line testing during manufacture and high-throughput testing during development. However, samples are often turbid, multiply scattering light and preventing the direct use of common sizing techniques. Differential dynamic microscopy (DDM) is a powerful technique for analysing video microscopy of such samples, measuring diffusion and hence particle size without the need to resolve individual particles while free of substantial user input. However, when applying DDM to a flowing sample, diffusive dynamics are rapidly dominated by flow effects, preventing particle sizing. Here, we develop ``flow-DDM'', a novel analysis scheme that combines optimised imaging conditions, a drift-velocity correction and modelling of the impact of flow. Flow-DDM allows a decoupling of flow from diffusive motion that facilitates successful particle size measurements at flow speeds an order of magnitude higher than for DDM. We demonstrate the generality of the technique by applying flow-DDM to two separate microscopy methods and flow geometries.}
\end{tabular}
\end{@twocolumnfalse}
\vspace{0.6cm}
]

\renewcommand*\rmdefault{bch}\normalfont\upshape
\rmfamily
\section*{}
\vspace{-1cm}


\footnotetext{\textit{SUPA and School of Physics and Astronomy, University of Edinburgh, King's Buildings, Edinburgh EH9 3FD, United Kingdom. E-mail: james.a.richards@ed.ac.uk, \mbox{vincent.martinez@ed.ac.uk}, j.arlt@ed.ac.uk}}
\footnotetext{\dag~Electronic Supplementary Information (ESI) available: containing details on the impact of the form of the residual velocity distribution, depth dependence, $q$-dependent fitting, extracted velocity distributions and far-field correlator results.}

\section{Introduction}

Solid micron-sized particles, say from \SI{100}{\nano\metre} to several \si{\micro\metre}, dispersed in a liquid are ever present in our lives. These colloidal suspensions form the basis of consumer formulations (\textit{e.g.} sunscreen), construction materials, and even pharmaceuticals or food. In all these applications the particle size can be of critical importance for performance, controlling the strength of concrete\cite{bentz1999effects} and paint film formation,\cite{kan1999role} or the rates of drug adsorbtion.\cite{sandri2014role} Particle size can even influence our sensory perception of materials, as with the taste of chocolate.\cite{ziegler2001role}

Measuring the size of particles in formulations is therefore an important task, both during development (\textit{e.g.} high-throughput testing), but also in real-time during manufacture to ensure a consistent formulation. To achieve these goals it is necessary to characterise a suspension not just in a quiescent (non-flowing) state but also under flowing conditions. For quiescent samples, various approaches to particle sizing exist, for which the reference method is to determine size directly from high-resolution electron microscopy images.\cite{bell2012emerging} However, this requires dry particles (it is not an \textit{in situ} method) and expensive equipment. More routine laboratory techniques for sizing particles in suspension include the well-established methods of static and dynamic light-scattering (SLS and DLS).\cite{berne2000dynamic} SLS measures the particle form factor (and hence size) from the average intensity scattered; in contrast, DLS measures the free diffusion coefficient, $D_0$, via temporal fluctuations of the scattered intensity due to Brownian motion. From $D_0$, the particle diameter, $d$, can be extracted via the Stokes-Einstein relation. DLS has been extended to flowing systems for in-line testing, but the measured particle size is impacted by flow speed.\cite{leung2006particle}

However, for formulation science a more fundamental issue arises for both SLS and DLS, as the techniques are strongly affected by multiple scattering, where photons interact with more than one particle before reaching the detector. Although suppression of multiple scattering is possible using advanced DLS techniques,\cite{pusey1999suppression,urban1998characterization} highly dilute and transparent samples are required for standard commercial DLS setups. For formulations, which may even be turbid, this is an excessively restrictive requirement.

This limitation arises from the fact that DLS operates on a large scattering volume. One can also extract size from dynamics in a smaller volume by tracking individual particles from video microscopy.\cite{finder2004analysis}
However, this approach requires identifying individual particles, a task which becomes impracticable for smaller particles ($d \lesssim \SI{500}{\nano\metre}$) or in non-dilute, turbid systems,\cite{crocker1996methods} although one which machine learning is being applied to.\cite{newby2018convolutional}
Using differential dynamic microscopy (DDM) \cite{trappe2008differential}, a digital Fourier analysis of video microscopy, we avoid both user inputs and particle location.
DDM has been used to characterise the micro-rheological properties of fluids;\cite{edera2017differential, bayles2017probe, sanchez2018microliter} to enable high-throughput measurements of micro-organism motility;\cite{wilson2011differential, martinez2012differential, jepson2019high} and to measure particle diffusion in complex environments,\cite{latreille2019spontaneous,regan2019bridging} under external fields,\cite{reufer2012differential} and even in dense or turbid systems.\cite{lu2012characterizing, lazaro2019glassy, pal2020anisotropic}

However, for flowing suspensions the fluid's velocity can impact many particle-sizing techniques, causing an apparent increase in diffusion and an underestimation of particle size.\cite{leung2006particle, tong2016flow} Therefore, for reliable particle sizing microscopic diffusive motion must be disentangled from the impact of bulk flow. The effect of flow on another digital Fourier microscopy technique\cite{philippe2016efficient} related to DDM has recently been suggested, but this was limited to exploring qualitative changes in the microscopic dynamics of soft solids.\cite{aime2019probing}

Here, we present ``flow-DDM'', a novel DDM-based analysis scheme to quantitatively measure diffusive dynamics in flowing samples using a combination of drift-velocity correction and an appropriate theoretical model. Respectively, these reduce the contribution of the flow to the dynamics and allow a careful decoupling of the diffusive dynamics from the residual effects of flow. Using dilute colloidal suspensions, we systematically validate flow-DDM as a function of flow speed for the accurate measurement of particle size. We find that flow-DDM outperforms current DDM techniques by an order of magnitude in the maximum possible flow speed. We establish a measurement protocol, bounds for reliable diffusion measurements and a guide to optimise the imaging method, which together could be widely applied for particle sizing in a multitude of flowing samples. This is demonstrated using phase-contrast microscopy of Poiseuille flow and fluorescence confocal microscopy of a rheometric shear flow.

\section{Modelling the impact of flow \label{sec:model}}

\subsection{Differential Dynamic Microscopy}

Before describing the impact of flow, we shall briefly cover how DDM can be used for particle sizing in a quiescent sample. DDM characterises the spatio-temporal density fluctuations within a sample by analysing microscopy movies, $I(\vec r, t)$, of a sample region. Specifically, one computes the differential intensity correlation function (DICF), also known as the image structure function:
\begin{equation}
    g(\vec q,\tau) = \left \langle |\tilde I (\vec q, t + \tau) - \tilde I(\vec q, t)|^2 \right \rangle_t\\
    \label{eq:DICF}
\end{equation}
with $\tilde I(\vec q, t)$ the Fourier transform of $I(\vec r, t)$ and $\tau$ the delay time. Under appropriate imaging conditions and assuming the intensity fluctuations are proportional to fluctuations in sample density ($\Delta I \propto \Delta \rho$) the DICF can be written as \cite{reufer2012differential}
\begin{equation}
    g(\vec q,\tau) = A(\vec q)[1-\Re (f(\vec q, \tau))] + B(\vec q),
    \label{eq:DICF-general}
\end{equation}
where $A(\vec q)$ characterises the signal amplitude (which will depend both on sample properties, such as the particle's form factor, and the imaging system) and $B(\vec q)$ accounts for uncorrelated background noise. Here $f(\vec{q},\tau)$, often known as the intermediate scattering function (ISF), is the $\vec q$-Fourier component of the probability of the particle displacements, $\delta \vec r = \vec r_j(t+\tau)-\vec r_j(t)$,
\begin{equation}
    f(\vec q,\tau) = \langle e^{i\vec q \cdot\delta \vec r} \rangle_{j,~t}\,,
    \label{eq:ISF_explicit}
\end{equation}
with brackets denoting averages over all particles $j$ and time $t$. In the absence of net flow, $f(\vec q, \tau)$ is a real valued function and if the underlying dynamics are isotropic it only depends on $|\vec q|=q$, leading back to the more familiar, simplified expression: $g(q) = A(q) [1-f(q,\tau)] + B(q)$.
To extract information from the DICF, a parameterised ISF must be fitted. For non-interacting Brownian particles with diameter, $d$,
\begin{equation}
    f(\vec q, \tau) = f_{D}(|\vec q|=q,\tau) = e^{-D q^2 \tau},~
    D = {k_B T}/{(3\pi\eta_s d)}.
    \label{eq:diff}
\end{equation}
with $k_B T$ the thermal energy, $\eta_s$ the solvent viscosity and $D$ the extracted diffusivity. However, flow brings anisotropy in particle displacement and complexity to the ISF: to size particles we must disentangle microscopic dynamics and macroscopic flow.

\subsection{Impact of flow on DDM}
Under flowing conditions, the total displacement of a Brownian particle, $\delta \vec r$, is the sum of diffusive motion and ballistic motion due to flow, $\delta \vec r_v$. Using Eq.~\ref{eq:ISF_explicit}, the ISF can be expressed as a product of separate processes:\cite{aime2019probing}
\begin{equation}
    f(\vec q,\tau) = \prod_i f_i(\vec q, \tau) = f_D \cdot f_v \cdot f_\mathrm{FS}, 
\end{equation}
which includes contributions from diffusive motion ($f_D$), flow related motion ($f_v$) and finite size effects ($f_\mathrm{FS}$). As the total ISF is a \emph{product}, whenever a single component $f_i \rightarrow 0$, the total ISF $f \rightarrow 0$. Therefore, the fastest decorrelation process will dominate the entire response, leaving slower processes immeasurable. This means that, to measure particle size diffusion must be the fastest decorrelation process and that we must then understand the detailed impact of flow on the ISF.

\begin{figure}[tb]
    \centering
    \includegraphics{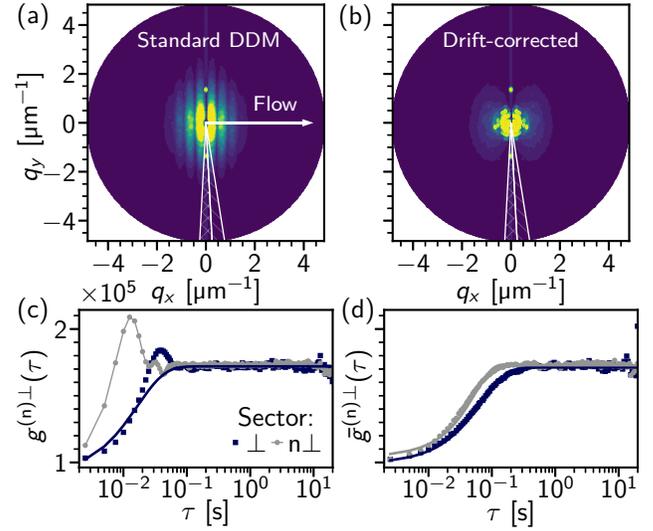}
    \caption[Impact of shear on DICF]{Impact of shear and drift correction on DICF (a)~DICF for DDM correlator, $g(\vec q)$ (Eq.~\ref{eq:DICF}) at delay time $\tau =$ \SI{0.02}{\second}, for Poiseuille flow at \SI{20}{\micro\litre\per\minute} and \SI{500}{\micro\metre} imaging depth ($\V = \SI{630}{\micro\metre\per\second}$). Colour map: light, high $g$ values, and dark, low $g$ values, flow direction indicated by arrow. Perpendicular ($\perp$, cross-hatched) and near-perpendicular (n$\perp$, hatched) sectors used to define $g ^{({\rm n})\perp}(q)$ with half-width $\theta = \SI{3}{\degree}$. (b)~DICF after drift correction, $\bar g$ (Eq.~\ref{eq:correlator}), colour scale unchanged. (c)~Time dependence of non-corrected DICF in (a) at $q = \SI{3}{\per\micro\metre}$, $g(\tau)$. Symbols: dark (blue), $\perp$; and, light (grey), n$\perp$. Line, anisotropic-DDM, diffusive fit of $g^\perp$ ($D = \SI{6.8}{\micro\metre^2\per\second}$). (d)~Drift-corrected DICF from (b): symbols, as in (c); lines, flow-DDM, Eq.~\ref{eq:f_product} ($D = \SI{1.55}{\micro\metre^2\per\second}$, $\Delta v = \SI{88}{\micro\metre\per\second}$).}
    \label{fig:ISF}
\end{figure}

A uniform steady flow, with velocity $\vec v$, will shift the position of each particle by $\delta\vec r_v= \vec v \tau$ in addition to diffusive motion, introducing a phase shift into the ISF:
\begin{equation}
    f_v(\vec q,\tau) =  e^{i\vec q \cdot \vec v \tau}, \;\mathrm{thus}\; \Re \{f_v(\vec q,\tau)\}= \cos(\vec q \cdot \vec v \tau).
    \label{eq:f_v}
\end{equation}
This is apparent in the DICF as `waves' in the direction of flow, as illustrated in Fig.~\ref{fig:ISF}(a), which shows a typical experimental DICF, in the $(q_x,q_y)$ plane at one delay time $\tau = \SI{0.02}{\second}$, obtained for Brownian particles flowing with mean velocity $\V = \SI{630}{\micro\metre\per\second}$ (see Sec.~\ref{sec:method_cap} for experimental details). Equation~\ref{eq:f_v} implies that flow should not contribute to the DICF in the direction perpendicular to the flow, $g^{\perp}$, as $\vec q \cdot \vec v = 0$, and $f_v^{\perp} = 1$. However, as images are composed of finite-sized pixels, measurements of $g^{\perp}$ require averaging over a finite-size sector with half-width $\theta$ and thus $\vec q$ is only \emph{approximately} perpendicular, Fig.~\ref{fig:ISF}(a)~(hatched). In practice, we found a minimum of $\theta \approx \SI{3}{\degree}$ is required to obtain measurable $g^{\perp}$ from a 256 pixel image. Therefore, this sector still contains a velocity component ($\sim \theta |\vec v|$), which introduces a decorrelation timescale ($t_v \sim 1/q\theta |\vec v|$), that for even moderate flow velocities can dominate over diffusion ($t_v \ll t_D = 1/Dq^2$). This velocity component leads to a non-monotonic (and assuredly non-diffusive) $g^{\perp}(\vec q, \tau)$ set by $f_v$ rather than $f_D$, Fig.~\ref{fig:ISF}(c)~[(blue) squares]. The non-monotonic behaviour is exacerbated in the adjacent sector~[(grey) circles]. 
We refer to a simple diffusive fit to $g^{\perp}$ as ``anisotropic-DDM'', a technique which has been used for particles influenced by a magnetic field.\cite{reufer2012differential,pal2020anisotropic}

A combination of a finite field of view and flow will also cause decorrelation due to particles leaving the image (and being replaced by on average uncorrelated particles).\cite{aime2019probing} This introduces a finite-size term into the total ISF, which for flow along the $x$ direction takes the form 
\begin{equation}
    f_\mathrm{FS} = \max\left[(1- |v_x| \tau/L_x),0\right],
    \label{eq:f_finite}
\end{equation}
where $L_x$ is the image size in the flow direction. This sets a hard upper limit for DDM-based measurements under flow, as the diffusive dynamics must lead to decorrelation on a timescale faster than $L_x/v_x$, whereupon particles disappear from the field of view.

\subsection{Flow-DDM}
As stated, anisotropic-DDM is quickly overwhelmed by the presence of flow and the remaining velocity component. Here, we present a new analysis scheme, flow-DDM, that allows a reduction of the flow contribution and a decoupling of the diffusive motion of Brownian particles from the background flow. Conceptually, the effect of flow on a system moving with a well defined uniform speed, $\vec v$, can be minimized by simply observing its dynamics in a co-moving frame of reference. Recording movies directly in a co-moving frame of reference is obviously challenging, but by determining the mean drift speed $\langle \vec{v} \rangle$ in the laboratory frame of reference it is then straightforward to shift the images when computing the DICF. The resulting ``drift-corrected DICF'' can then be fitted with an appropriate model that takes into account diffusive motion and the fact that in most practical scenarios there will be a spread in flow speeds.

\subsubsection{Drift-corrected DICF}
We first need to measure the mean flow velocity, for which several methods exist such as particle tracking velocimetry or particle imaging velocimetry.\cite{besseling2009flow} However, the recently introduced method of phase dynamic microscopy\cite{colin2014fast} ($\varphi$DM) is particularly suitable in the current context because it is a digital Fourier method that does not require particle resolution and can be readily integrated with DDM. At high flow speeds the dominant change between frames is the translation, which in Fourier space leads to a cumulative phase shift $\varphi(\vec q) = \vec q  \cdot \vec v \tau$ (from Eq.~\ref{eq:f_v}). The drift velocity $ \vec{v} $ can then be estimated from the gradient of $\varphi$; by averaging over a sufficiently long movie segment. The method has been shown to work over a wide range of speeds, even when the displacements due to random motion start to dominate.\cite{colin2014fast}

Having measured the mean flow velocity, $\langle \vec{v} \rangle$, we can then compute the drift-corrected DICF:
\begin{equation}
    \begin{split}
        \bar g(\vec q,\tau) &= \left \langle |\tilde I (\vec q, t + \tau)e^{-i\vec q \cdot \langle \vec{v} \rangle \tau} - \tilde I(\vec q, t)|^2 \right \rangle_t\\
        &= A(\vec q)\,[1-\bar f(\vec q,\tau)] + B(\vec q).
    \end{split}
    \label{eq:correlator}
\end{equation}
Equation~\ref{eq:correlator} allows reduction of the flow contribution, as both the 'waves' and non-monotonic behaviour of the DICF [Fig.~\ref{fig:ISF}(a) and (c)] are not apparent in the drift-corrected DICF [(b) and (d)]. However, we note that the drift-corrected DICF is clearly not radially symmetric, indicating that there is still some residual contribution from the flow. This is due to the fact that there is actually a \emph{distribution} of flow speeds about the mean. This speed distribution must be considered to allow accurate measurements of particle size at high flow speeds.

\begin{figure}[tb]
    \centering
    \includegraphics{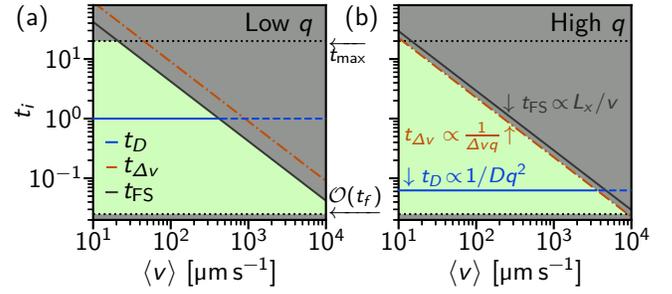}
    \caption{Decorrelation time `phase diagram'. (a) ISF decorrelation timescales, $t_i$, as a function of average mean drift velocity, $\langle v \rangle$, at low wavevector, $q = \SI{1}{\per\micro\metre}$. Lines: blue, diffusion time ($t_{D} = Dq^2$, for $D = \SI{1}{\micro\metre^2\per\second}$, solid when measurable, dashed otherwise); orange (dot-dashed) velocity distribution [$t_{\Delta v} = 4.5/(\Delta v q \theta)$, $\Delta v = 0.1 \V$]; dark grey (solid), finite-size effect ($t_\mathrm{FS} = 0.63\, L_x / \V$, $L_x = \SI{665}{\micro\metre}$); and dotted lines, standard DDM limits [lower, frame time limit ($\sim 10 t_f$); and upper, movie length ($t_{\max}$)]. Shading: light green, diffusion measurable; grey, decorrelation before diffusion. (b)~Equivalent decorrelation time diagram at high $q = \SI{3}{\per\micro\metre}$, sharing $t_i$ axis.}
    \label{fig:phase}
\end{figure}

\subsubsection{Modelling of the drift-corrected ISF}

To account for the residual effects of flow the drift-corrected ISF remains a product of three contributions,
\begin{equation}
    \bar f(\vec q,\tau) = \prod_i f_i(\vec q, \tau) = f_D \cdot f_{\Delta v} \cdot f_\mathrm{FS},
    \label{eq:f_product}
\end{equation}
but now including $f_{\Delta v}$ to account for the residual velocity distribution.
Such distributions in the flow velocity originate from several causes. Indeed, as the sample will be flowing through a geometry with fixed boundaries, there must be a velocity gradient (or shear).
As we image a finite volume due to the depth of field, this causes a range of particle speeds to be captured.
Additionally, there is often a small spatial variation of velocities across the field of view, for example, due to the speed profile in Poiseuille flow, or the flow speed may even vary over time. In all these cases, after correcting for the average velocity there will be a distribution of residual velocities, $P(\Delta \vec v^{\prime})$, which we characterise by the width $\Delta v$. In the following we assume that this residual motion is purely in the direction of the original flow and we now drop vector notation for velocities (see ESI$^{\dag}$ Section~S1 for comments on more general residual motion).

Note that the width of this distribution is in principle not set by the mean speed alone, \textit{e.g.} in rheometric cone-plate flow the shear rate is fixed (fixing the velocity distribution for a given optical section) but the speed varies with height.
However, it is important to realise that in practice for a given imaging region the velocity distribution will increase with the mean speed in a linear fashion, $\Delta v = k \langle v \rangle$, with the proportionally constant dependent on the flow geometry, but assumed to be less than 1 (for imaging away from the boundaries). 

To size particles, we first restrict our analysis to the perpendicular sector, ${f}_{\Delta v}^{\perp}$, for which the impact of ${\Delta v}$ is minimised (just as with $\V$ for $f^{\perp}_v$). This attempts to ensure that diffusion causes decorrelation in Eq.~\ref{eq:f_product}. For tractability, we assume a uniform distribution of residual velocities (-$\Delta v$ to +$\Delta v$) and use a small angle approximation for the phase shift, \textit{i.e.}~$\vec q \cdot \Delta \vec v^{\prime} \tau \approx q \Delta v^{\prime} \theta^{\prime} \tau$. Integration over the residual velocity distribution $(\Delta v^{\prime})$ and then sector angle ($\theta^{\prime}$) thus yields:
\begin{equation}
\begin{split}
    {f}_{\Delta v}^{\perp}(q,\tau) &\propto \int_{-\theta}^{+\theta} \int_{-\infty}^{+\infty} P(\Delta v^{\prime}) \exp(-i q \theta^{\prime} \Delta v^{\prime} \tau) {\rm d}\Delta v^{\prime}{\rm d}\theta^{\prime} \\
    &= \Si(q \Delta v \tau \theta)/(q\Delta v \tau \theta),
\end{split}
\label{eq:f_shear}
\end{equation}
where $\Si$ is the sine integral and proportionality such that $f(\tau \to 0) = 1$. Note that although the assumption of a uniform $P(\Delta\!v^{\prime})$ is evidently an idealisation, it is sufficient to capture the features of more realistic distributions within the frame work of our flow-DDM protocol (see ESI$^{\dag}$ Section~S1).

\subsubsection{Optimisation of flow-DDM\label{sec:opti}}

To see how best to extract an accurate particle size from $\bar g^{\perp}$ over the greatest possible range of flow speeds we must consider relative decorrelation times for different components of the ISF, where $f_i(t_i) = 1/e$ in Eq.~\ref{eq:f_product}.
The decorrelation time for diffusion, $t_D = 1/Dq^2$, does not depend on $\V$, Fig.~\ref{fig:phase}~[(blue) solid line], but it does decrease strongly with increasing $q$.

Finite-size effects by contrast lead to $t_{\mathrm{FS}}= 0.63\, L_x/v_x$, independent of $q$.
Therefore decorrelation is predominantly due to diffusion for speeds up to $v_x \approx 0.63\, q^2 D L_x$, \textit{i.e.}~this effect becomes less important at higher $q$, \textit{cf.}~Fig.~\ref{fig:phase}(a) blue and dark grey lines.
By acquiring images with a large field of view $L_x$ and high spatial resolution (to access high $q$) finite size effects can be be greatly reduced. 
But the faster dynamics at higher $q$ also require high frame rates, which in modern scientific cameras and confocal laser scanning systems decreases with the height $L_y$ of the image.  
In practice, these requirements are most effectively achieved by taking a rectangular image, with the long axis of the field of view aligned with the flow direction: we use $L_x = 4L_y$ throughout.

The decorrelation time caused by the distribution of speeds, from Eq.~\ref{eq:f_shear}, 
\begin{equation}
    t_{\Delta v} =\frac{4.5}{q\theta \Delta v}
    \label{eq:tDV}
\end{equation}
decreases with both the width of the speed distribution (and thus flow speed) and with $q$.
Therefore, we can again increase the impact of diffusion, this time relative to $\Delta v$, by looking at higher $q$, [\textit{cf.}~(orange) dot-dashed lines, Fig.~\ref{fig:phase}(a) and (b), where we take $\Delta v = 0.1 \langle v \rangle$], and hence measure particle size at higher speeds. Experimentally, we access high $q$ using relatively high magnifications. This has the added benefit of reducing the imaged width, and therefore the contribution to $\Delta v$ from velocity variation in the $y$ direction. For some microscopy methods, \textit{e.g.}, brightfield, the depth of field is also decreased at high $q$\cite{giavazzi2009scattering}, reducing any contribution to $\Delta \! v$ from the velocity gradient in $z$. However, the maximum useful magnification is limited by the drop in signal amplitude [$A(q) \ll B(q)$], as without accessing higher $q$ greater magnification only increases finite size effects.

\subsubsection{Consistency check}

While we have now maximised the impact of diffusion relative to the flow on decorrelation, we must also discriminate between the two processes to determine the reliability of the measurement.
For finite-size effects we can estimate $t_\mathrm{FS}$ independently from $\V$; but, in the perpendicular sector there is no robust way to discriminate between $\Delta v$ and $D$ over a limited $q$ range, as both $f_{\Delta v}$ and $f_D$ decrease monotonically. However, diffusion is isotropic, while the impact of shear depends on angle. We therefore consider a sector that is adjacent to the perpendicular sector, $\bar g^{\rm n \perp}$ [Fig.~\ref{fig:ISF}(a)], with
\begin{equation}
\begin{split}
    {f}_{\Delta v}^{\rm n \perp}(q,\tau)  &\propto \int_{\theta}^{3\theta} \int_{-\infty}^{+\infty} P(\Delta v^{\prime}) \exp(-i q \theta^{\prime} \Delta v^{\prime} \tau) {\rm d}\Delta v^{\prime}{\rm d}\theta^{\prime} \\
    &=\left[\Si(3 q \Delta v \tau \theta) - \Si(q\Delta v \tau \theta )\right]/(2 q\Delta v \tau \theta).
\end{split}
    \label{eq:f_shearNear}
\end{equation}
Decorrelation due to $\Delta v$ now occurs at a more rapid rate ($\sim 3\times$ compared to $f^{\perp}_{\Delta v}$) and we can separately probe $\Delta v$ by simultaneously fitting two sectors of the DICF and establish whether the measured particle size is reliable, \textit{i.e.}~$t_D \ll \{t_\mathrm{FS},~t_{\Delta v}\}$. This combination of drift correction, imaging optimisation and fitting together we term ``flow-DDM''.

\section{Experimental materials and methods\label{sec:methods}}

\begin{figure}[tb!]
    \centering
    \includegraphics{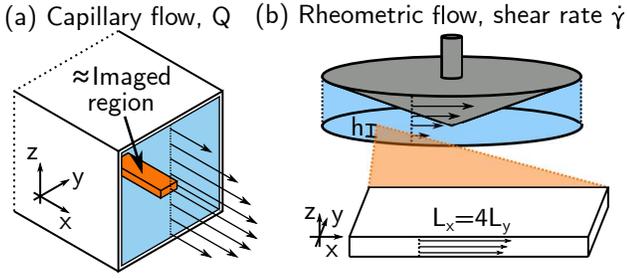}
    \caption{Imaging and flow geometries. (a)~Capillary flow, \SI{1}{\milli\metre} square with flow rate $Q$. Arrows indicate velocity direction and spatial variation; example imaging region shown in orange at height $z$ with flow direction $x$. (b)~Rheometric flow with confocal microscopy. Flow (arrows) generated by rotating cone (grey) above glass coverslip with shear rate, $\dot\gamma$ (velocity gradient, d$v$/d$z$). Imaging region highlighted with 4:1 aspect ratio ($L_x$:$L_y$) aligned with flow direction.}
    \label{fig:geometry}
\end{figure}
We now turn to look at applying flow-DDM to measure particle size for a dilute colloidal suspension and demonstrate it using two different microscopy techniques and flow geometries.

\subsection{Poiseuille flow\label{sec:method_cap}}

First, we use a dilute ($\phi = 0.004\%$) suspension of spherical polystyrene particles in water, with a nominal diameter of \SI{300}{\nano\metre} flowing through a \SI{1}{\milli\metre} square capillary with a controlled flow rate, giving Poiseuille flow, Fig.~\ref{fig:geometry}(a). Images were taken using phase contrast microscopy (20x/0.5 objective at 400 frames per second for $t_{\max} = \SI{20}{\second}$). The rectangular images ($1024 \times 256$ pixels image, \SI{0.65}{\micro\metre}/px $\rightarrow$ \SI{166}{\micro\metre} by \SI{665}{\micro\metre}) are aligned along the centre of the capillary. At a given flow rate, $Q$, the capillary is then imaged at multiple focal depths, $z$.

To establish a reference diffusion coefficient, \textit{i.e.}~the free-diffusion coefficient ($D_0$), quiescent samples were recorded in the same conditions. Using standard DDM (Eqs.~\ref{eq:DICF} and \ref{eq:diff}), a $q$-dependent diffusion coefficient was extracted, Fig.~\ref{fig:rest}(a). The diffusivity, $D_0 = \langle D(q) \rangle = \SI{1.52(1)}{\micro\metre^2\per\second}$ (averaging over $q = 1.0$ to \SI{3.0}{\per \micro\metre}) implies a particle size of $d=\SI{298(3)}{\nano\metre}$ at \SI{22}{\celsius}. 

\subsection{Rheo-confocal flow\label{sec:method_con}}

To explore the general application of flow-DDM to other microscopy techniques and flow geometries, we use a confocal microscope coupled to a rotational stress-controlled rheometer\cite{besseling2007three} (Anton Paar MCR 301), Fig.~\ref{fig:geometry}(b). Images were taken using an inverted confocal laser-scanning microscope [Leica SP8, 20x/0.75 objective)], a technique previously used with DDM to measure dense quiescent systems.\cite{lu2012characterizing} The sample is a dilute ($\phi = 0.5\%$) suspension of fluorescently-dyed poly(methyl methacrylate) particles stabilised with poly(vinyl pyrrolidone); the particles are suspended in a density matched 21~wt.\% caesium chloride solution to prevent sedimentation and screen electrostatic interactions. Images are taken at 50 frames per second for $t_{\max}= \SI{200}{\second}$ with a $1024\times256$ resolution and \SI{0.455}{\micro\metre} pixel size ($\SI{466}{\micro\metre}\times\SI{116}{\micro\metre}$ field of view).

\begin{figure}[tb]
    \centering
    \includegraphics{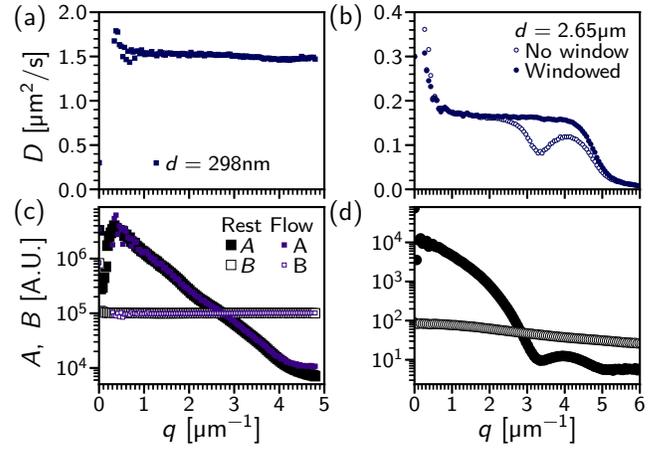}
    \caption[Quiescent samples]{Diffusion measurements of quiescent samples. (a)~Phase contrast microscopy of a dilute suspension of \SI{300}{\nano\metre} polystyrene particles as a function of wavevector, $q$. (b)~Confocal microscopy of a dilute colloidal suspension, poly(methyl methacrylate) in CsCl solution, $\sim \SI{2}{\micro\metre}$. Symbols, $D(q)$ for: filled (blue) squares, standard DDM protocol; and, open squares, Hanning-windowed data. (c)~Signal [filled, $A(q)$] and noise [open, $B(q)$] for $D$ measurements in (a). Large symbols at rest, small symbols under flow at $\V = \SI{100}{\micro\metre\per\second}$. (d)~$A(q)$ and $B(q)$ for $D$ in (b).}
    \label{fig:rest}
\end{figure}

In the quiescent state, a plateau in $D(q)$ is seen for $q = 1$ to \SI{2.5}{\per\micro\metre}, Fig.~\ref{fig:rest}(b)~(open circles). Due to the small image width ($L_y$) used for flow-DDM,  ``spectral leakage'' leads to an apparent drop in diffusivity: at high $q$ values, corresponding to length scales smaller than the particle, $g(q,\tau)$ is distorted due to particles cut off at the image boundaries.\cite{giavazzi2017image} This effect is mitigated by smoothing the image boundaries using a Hanning window, (\textit{cf.}~open and filled symbols); all further diffusion measurements presented are from windowed images. Averaging $D(q)$ from \SIrange{1.0}{3.0}{\per\micro\metre}, gives $D_0 =  \SI{0.164(1)}{\micro\metre^2\per \second}$ and a particle diameter of \SI{2.65(1)}{\micro\metre} at \SI{20}{\celsius}. At low $q$ there is an apparent rise in $D$ due to diffusion out of the optical section.\cite{giavazzi2017image}
Additionally, we can also estimate the particle size from $A(q)$, Fig.~\ref{fig:rest}. Considering high resolution fluorescence imaging of a dilute suspension, we expect $A(q)$ to be proportional to the particle form factor, for which a first minimum should occur at $qd/2\approx 4.5$ by considering the Fourier transform of a uniform intensity and neglecting the point spread function. The minimum at $q =  \SI{3.4}{\micro\metre}^{-1}$, Fig.~\ref{fig:rest}(d), results in an estimated diameter $d\approx \SI{2.64}{\micro\metre}$, in quantitative agreement with results from the measured $D_0$.

To create flow, a $\SI{1}{\degree}$, $\SI{50}{\milli\metre}$ diameter cone-plate geometry generates a uniform shear rate, $\dot\gamma$, with the velocity gradient perpendicular to the imaging plane. The shear rate is set by the rotational speed of the rheometer. Imaging at an increasing depth into the sample, $h = 10$ and $\SI{20}{\micro\metre}$, increases the translational speed $\V  = \dot \gamma h$; greater depths could not be used due to high sample turbidity. Images are taken at a radius of $\approx \SI{20}{\milli\metre}$ from center of the cone, to ensure the direction of the rotational flow does not vary significantly along the flow direction, $x$.

\section{Results and discussion}

\subsection{Poiseuille flow\label{sec:cap_results}}

\begin{figure}[tb!]
    \centering
    \includegraphics{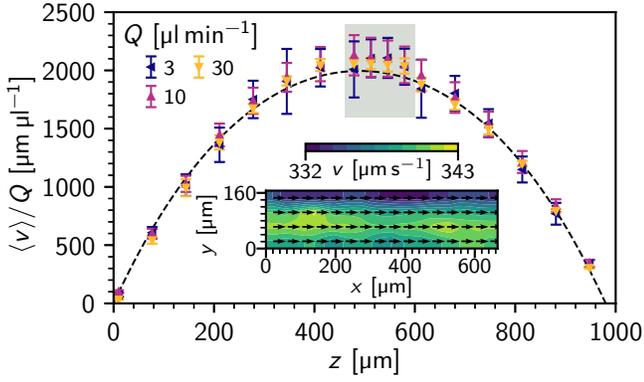}
    \caption{Velocity variation in a capillary. (a)~Average drift velocity, $\V$, as a function of imaging depth, $z$, collapsed by flow rate, $Q$. Symbols: varying $Q$, see inset legend, error bars indicate standard deviation in $v(t)$ from 100 frame (\SI{0.25}{\second}) subsections; dashed line, expected $\V$ from \SI{0.98}{\milli\metre} square capillary flow profile, averaged over \SI{166}{\micro\metre} image width. Grey shading, positions used for particle sizing measurements. Inset: spatial variation of velocity in the centre of the channel at $Q = \SI{10}{\micro\litre\per\minute}$, shade (color) indicates velocity (see scale above) and arrows direction. Average drift velocity is extracted from \SI{40}{\micro\metre} sub-regions and linearly interpolated.}
    \label{fig:cap_velocity}
\end{figure}

We now establish the effectiveness of flow-DDM and investigate the limiting factors for reliable measurement. We measured diffusivity of a dilute colloidal suspension with increasing flow rate through a capillary, which we compare to the free-diffusion coefficient $D_0$ obtained from quiescent conditions. However, the flow velocity in a capillary varies strongly with position. We show in Fig.~\ref{fig:cap_velocity} the average flow velocity $\V$, measured in the $(x,y)$-plane center of the capillary using $\varphi$DM and normalised to the flow rate $Q$, as a function of the height of the focal plane ($z$) for several $Q$ values. We find a near parabolic flow profile, with the velocity reaching a maximum in the centre of the capillary, Fig.~\ref{fig:cap_velocity}~(symbols), matching the velocity predicted by Boussinesq\cite{boussinesq1868memoire} (dashed line). Temporal fluctuations in the flow speed may occur and would be included in error bars, but no systematic variation over $\sim t_{\max}$ was observed. Near the centre of the channel ($z = $~\SIrange{480}{580}{\micro\metre}), $\V$ is near constant and we therefore average over these four positions, although we present results across the full depth in Fig.~S2, ESI$^{\dag}$. These measurements are away from the top and bottom of the channel, where the strong gradient in $\V$ may combine with the optical section to produce a large $\Delta v$. As with $z$, there is also a velocity variation across the channel width, $y$. Measuring $\V$ in sub-regions of the image we can estimate this variation at $\approx 3\%$, with a \SI{11}{\micro\metre\per\second} spatial variation for $\V = \SI{338}{\micro\metre\per\second}$, Fig.~\ref{fig:cap_velocity}(inset).

From the measured $\V$, we computed the drift-corrected DICF for all positions (Fig.~S3 for typical $\bar{g}$, ESI$^{\dag}$). To extract a diffusion coefficient we simultaneously fit the perpendicular and near-perpendicular sectors of the drift-corrected DICF, Fig.~\ref{fig:ISF}(b), using Eqs.~\ref{eq:diff},~\ref{eq:correlator}--\ref{eq:f_shearNear} over a $q$ range of $\SI{3.0}{\per\micro\metre}$ to \SI{3.5}{\per\micro\metre} where $A/B>0.3$. $\V$ is taken as an input parameter, and \{$D(q)$, $\Delta v$, $A^{{\rm (n)}\perp}(q)$ and $B^{{\rm (n)}\perp}(q)$\} as the fitting parameters (Fig.~S3 for typical results as a function of $q$, ESI$^{\dag}$).

\begin{figure}[tb!]
    \centering
    \includegraphics{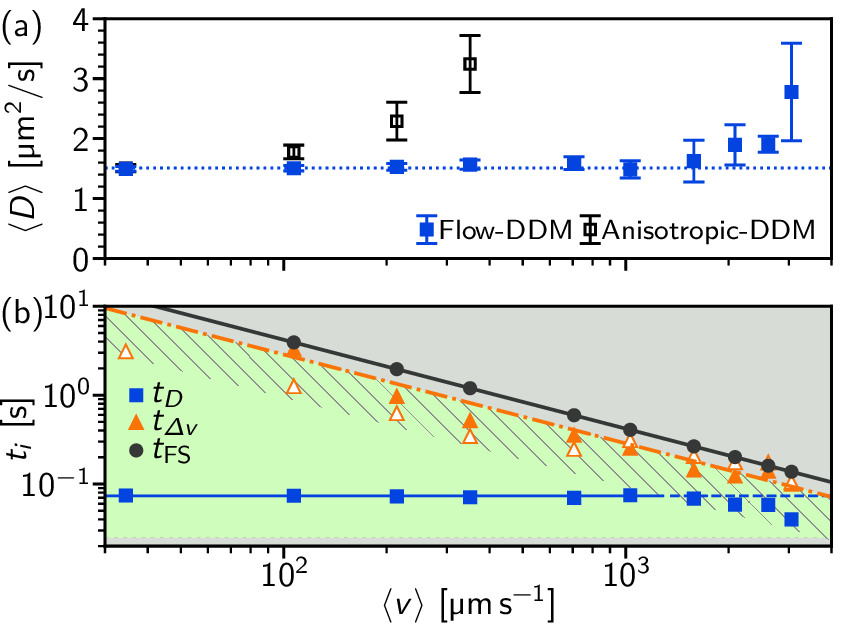}
    \caption{Measuring diffusion with varying capillary flow rate. (a)~Extracted diffusivities vs mean drift velocity, $\V$, averaging over 4 positions in channel centre. Symbols: filled (blue) squares, flow-DDM averaging $D(q)$ over $q = $~\SIrange{3.0}{3.5}{\per\micro\metre} with $\theta =$ \SI{3}{\degree} and open (black) squares, anisotropic-DDM. (b) Timescale phase diagram. Symbols, timescales at minimum $q$ used for flow-DDM, $q = \SI{3.0}{\per\micro\metre}$: (blue) squares, measured diffusion; solid (orange) triangles, extracted velocity distribution from flow-DDM; open (orange) triangles, velocity distribution from $v(t)$, Fig.~\ref{fig:cap_velocity}, using \SI{0.25}{\second} subsections and the difference between 5\textsuperscript{th} and 95\textsuperscript{th} percentiles; and (grey) circles, finite-size effect from $\V$. Lines and shading scheme as in Fig.~\ref{fig:phase}, with striped shading indicating factor three timescale separation.}
    \label{fig:capillary}
\end{figure}

We varied the flow rate in the range $Q = $~\SIrange{1}{90}{\micro\litre\per\minute}, resulting in nearly two decades of measured $\V$ in the middle region of the capillary (from \SIrange{34}{3000}{\micro\metre\per\second}).
Plotting the extracted diffusivity $\D$ against $\V$, Fig.~\ref{fig:capillary}(a), we find that $\D$ closely matches the quiescent measurement, $D_0$, up to $\SI{1000}{\micro\metre\per\second}$, \textit{cf.}~filled squares and dashed line.
Correspondingly, at the minimum $q$ used for averaging the diffusion timescale $t_D$ is far smaller than $t_{\Delta v}$ and $t_\mathrm{FS}$, Fig.~\ref{fig:capillary}(b), giving great confidence in the accuracy of the overall analysis, as discussed in Sec.~\ref{sec:opti}.
However, $t_D$ and $t_{\Delta v}$ become comparable at higher velocity $\V = \SI{1500}{\micro\metre\per\second}$ and so the error in $\D$ increases, before $\D$ itself increases at yet higher speeds.
For sizing, this would appear as a smaller particle.
Based on Fig.~\ref{fig:capillary}, we conclude that due to the present optimal imaging conditions $\Delta v$ is the limiting factor (as $t_{\Delta v} < t_{FS}$) and that $t_D \lessapprox t_{\Delta v}/3$ is necessary for reliable sizing measurements [Fig.~\ref{fig:capillary}(b) hatched region].
Using Eq.~\ref{eq:tDV}, this allows us to estimate the maximum velocity, $v_{\max} = \SI{1100}{\micro\metre\per\second}$, for reliable particle sizing by considering our measured $\Delta v\approx 0.1 v$ (Fig.~S4, ESI$^{\dag}$) and $\theta=\SI{3}{\degree}$.
Using larger $\theta = \SI{7.5}{\degree}$ sectors means $\Delta v$ will have a larger impact ($v_{\max} = \SI{430}{\micro\metre\per\second}$), and correspondingly we see a larger $\D$ measurement at a lower $\V \lesssim \SI{1000}{\micro\metre\per\second}$ (Fig.~S5, ESI$^{\dag}$).

A $\Delta v \sim 0.1 \V$ is larger than expected from variation across the width of the channel, Fig.~\ref{fig:cap_velocity}(inset).
It is instead related to temporal fluctuations, with $\Delta v$ measured with flow-DDM closely matched by the variation in $v(t)$, \textit{cf.}~Fig.~\ref{fig:capillary}(b) open and filled triangles.
The spatio-temporal velocity fluctuations mean that the contribution to $\Delta v$ from the optical section is insignificant, which results in consistent diffusivity measurements across the capillary, even as the velocity variation across the depth of field changes, see ESI$^{\dag}$ Section~2.
However, even if these measurements were not limited by flow stability, $t_\mathrm{FS}$ would soon impact measurements [Fig.~\ref{fig:capillary}(b), solid dark (grey) line], even with the rectangular field of view.

Comparing flow-DDM to existing DDM-based techniques, we see a significant improvement over anisotropic-DDM, \textit{i.e.}~using a perpendicular sector of $\theta=\SI{3}{\degree}$ and a simple diffusive fit (Eq.~\ref{eq:diff}) over the same $q$ range, Fig.~\ref{fig:capillary}(a). Flow-DDM enables reliable measurement of the free diffusion coefficient, $D_0$, and thus the particle size to $\V$ an order of magnitude faster than for anisotropic-DDM, for which $\D$ starts to significantly increase from $\V\lesssim \SI{100}{\micro\metre\per\second}$. The $\mathcal{O}(10)\times$ improvement is consistent with $\Delta v \sim 0.1 \V$ as the particle velocities are reduced 10 fold thanks to the drift-correction (Eq.~\ref{eq:correlator}).

Additionally, another recent technique based on DDM but using a higher-order ``far-field'' correlator has been suggested to eliminate the impact of translation due to flow (\textit{i.e.}~$\V$). This far-field correlator can be related to the \emph{magnitude} of the ISF, which should be translation invariant.\cite{aime2019probing} However, we find that even in quiescent conditions that interpretation of this correlator is challenging, as it yields a measured $D(q)$ lower than the expected $D_0$ (Fig.~S6, ESI$^{\dag}$), while for flowing samples the results vary proportionally with non--drift-corrected DDM (Fig.~S5, ESI$^{\dag}$). For quantitative results, we therefore use flow-DDM.

\subsection{Rheo-confocal flow\label{sec:rheo_results}}

\begin{figure}[tb]
    \centering
    \includegraphics{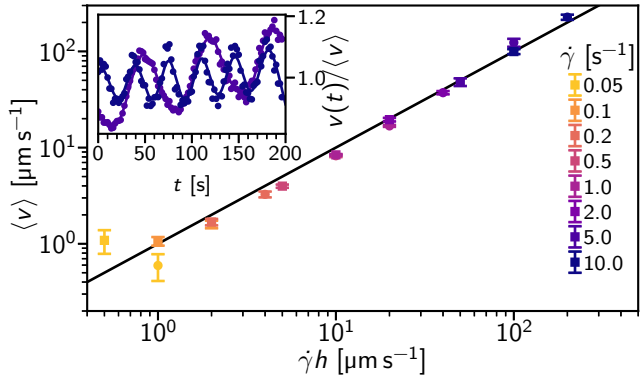}
    \caption[Rheometric flow velocity]{Rheo-confocal flow velocity. (a)~Extracted average drift velocity, $\V$, as a function of applied shear rate, $\dot{\gamma}$. Symbols, time averaged drift velocity extracted from phase shift between successive frames (error bars, standard deviation in $v$ extracted from \SI{2}{\second} subsections of movie; squares, imaging depth, $h = \SI{10}{\micro\metre}$; and circles, $h = \SI{20}{\micro\metre}$. Shear rate given by colour (or shade), see inset legend. Line, equality between $\V$ and nominal velocity, $\dot\gamma h$. Inset:~time-dependent drift velocity $v(t)$ for $h = \SI{10}{\micro\metre}$ at $\dot\gamma \geq \SI{5}{\per\second}$.}
    \label{fig:drift}
\end{figure}

We now demonstrate the general applicability of flow-DDM by using a setup with a different microscopy method, flow geometry and particle size. Here, we performed rheo-confocal imaging of micron-sized particles, and varied the flow velocity, $\V$, through the imaging height, $h= 10$ or \SI{20}{\micro\metre}, and applied shear rate, $\dot\gamma =$~\SIrange{0.05}{10}{\per\second}.
This setup allows us control of the mean speed independent of the velocity spread by imaging the flow within a well-defined optical section. 

Figure~\ref{fig:drift} shows $\V$ measured from $\varphi$DM (symbols) as a function of $\dot\gamma h$. The extracted average velocity closely matches the speed predicted for a shear flow, $\V = \dot\gamma h$ (line). However, at high shear rates ($\dot\gamma \geq 5$) there are noticeable oscillations in the flow speed (see inset), consistent with a slight geometry misalignment.\cite{dudgeon1993flow} The drift-corrected DICF, $\bar{g}$ (Eq.~\ref{eq:correlator}), was therefore calculated using a time-dependent drift velocity based upon a smoothed average of $\V$ from \SI{2}{\second} subsections, $v(t)$. We then fit $\bar{g}$ using the protocol developed for Poiseuille flow in Sec.~\ref{sec:cap_results}, but now using a $q$ range of \SIrange{2.0}{2.5}{\per\micro\metre} so that $A/B$ remains $\gtrsim 0.3$. The lower $q$ range consequently requires an increased $\theta$ of \SI{7.5}{\degree} to ensure an average over sufficient $\vec q$. Typical results for $\bar{g}$ and fits thereof are shown in Fig.~S3, ESI$^{\dag}$.

\begin{figure}[t]
    \centering
    \includegraphics{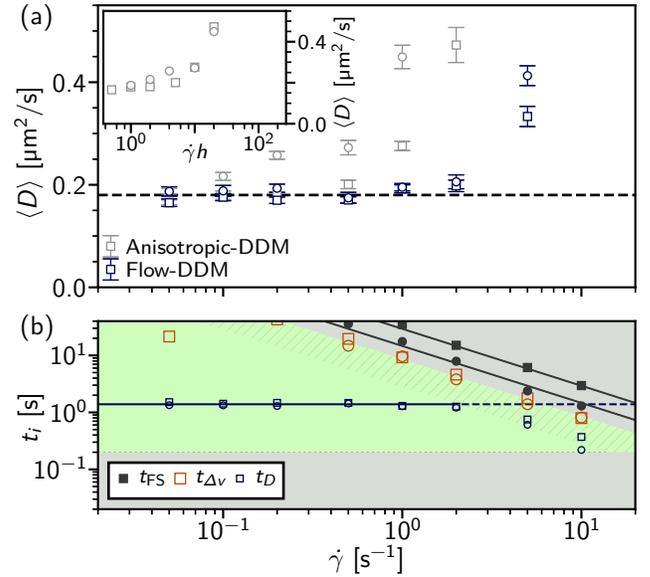}
    \caption[Measuring diffusion in rheometric flow]{Measuring diffusion in rheometric flow. (a)~Diffusion coefficient, $D$, as a function of applied shear rate, $\dot{\gamma}$, [$h = \SI{10}{\micro\metre}$ (squares) and $h = \SI{20}{\micro\metre}$ (circles)] averaging over $q = $~\SIrange{2.0}{2.5}{\per\micro\metre}. Symbols: dark (blue), $D$ from flow-DDM; and, light (grey) anisotropic-DDM. Inset: symbols, anisotropic-DDM vs nominal velocity ($\dot \gamma h$). (b)~Decorrelation times, $t_i$ at $q = \SI{2.0}{\per\micro\metre}$ for given terms from flow-DDM, symbols: small, $t_D$; large, $t_{\Delta v}$; and filled, $t_\mathrm{FS}$. Shading as in Fig.~\ref{fig:capillary}(b).}
    \label{fig:rheo}
\end{figure}

Figure~\ref{fig:rheo}(a) shows the measured $\D$ as a function of shear rate [light (blue) symbols]. At $\dot\gamma\lessapprox \SI{2}{\per\second}$, $\D\approx \SI{0.18}{\micro\metre^2\per\second}$, giving an inferred particle diameter of $d = \SI{2.4}{\micro\metre}$. The diffusivity is comparable to the rest measurement, $D_0=\SI{0.16}{\micro\metre^2\per\second}$, although there is an $\approx 10\%$ increase that may arise from a small change in the solvent viscosity due to temperature.

In order to understand the limits of flow-DDM we again need to compare the extracted decorrelation timescales shown in Fig.~\ref{fig:rheo}(b). 
First we should note that the decorrelation time associated with the spread in velocities, $t_{\Delta\!v}$, decreases with shear rate rather than the velocity: $t_{\Delta\!v}$ is the same for the two heights, $h$, presented here. 
This experimental data implies that $\Delta\! v = \Delta\! h \cdot \dot \gamma$, where we find $ \Delta\!h = \SI{2}{\micro\metre}$ (see ESI$^{\dag}$, Section~S4 for details). 
This lengthscale, $\Delta \! h$, is comparable to the quoted optical section of \SI{1.8}{\micro\metre} for our confocal imaging configuration, which suggests that $\Delta \!v$ arises from the velocity gradient across the depth of field in this shear flow. However, we cannot rule out a contribution from the time-dependent velocity as rapid changes may not be captured by the smooth interpolation of $v(t)$. Our optimised imaging settings ensured that finite size effects remain negligible, with $t_\mathrm{FS}$ the slowest of the three decorrelation processes, even at $h =\SI{20}{\micro\metre}$. So, diffusion (or size) measurements are limited by the increasing velocity distribution, with flow-DDM again producing reliable measurements for $t_D \lesssim t_{\Delta v}/3$, just as in Sec.~\ref{sec:cap_results}.

Using anisotropic-DDM, \textit{i.e.}~without drift correction, $\D$ shows an increase at much lower shear rates, Fig.~\ref{fig:rheo}(a)~(black symbols), and already increases at $\dot\gamma\gtrsim \SI{0.2}{\per\second}$ for $h=\SI{20}{\micro\metre}$ (circles). Here the rise in $\D$ occurs with the flow speed (see inset) rather than shear rate. The relative improvement seen for flow-DDM then depends on $h$, as the relevant velocity scale is changed from being set by the imaging depth ($\V = \dot\gamma h$) to being controlled by the effective optical section ($\Delta v = \dot\gamma \Delta\!h$): flow-DDM makes $(h/\Delta\!h) \times$ higher mean speeds accessible for size measurements.
Meanwhile, the far-field correlator again significantly underestimates diffusivity in quiescent conditions (Fig.~S6, ESI$^{\dag}$). Thus, flow-DDM appears as an exciting new technique to accurately measure free-diffusion and thus size particles under general flow conditions.

\section{Conclusions \label{sec:conc}}

In summary, we have proposed flow-DDM as a novel method to accurately measure free-diffusivity, and from this determine particle size, using microscopy videos of dilute suspensions of colloidal particles under flow. We have presented its theoretical framework and practical implementation for optimal measurements.

Flow-DDM is based on two main steps: 1)~computing the drift-corrected DICF, $\bar{g}$, from microscopy videos, which reduces the impact of flow onto the resulting experimental signal; and 2)~fitting $\bar{g}$ using an appropriate model of the particle motion (including diffusion, residual flow velocities and finite-size effects) coupled with an optimised fitting protocol that allows decoupling of the residual flow velocity distribution from the diffusive motion. We have validated flow-DDM using two different particle suspensions, demonstrating its general application by studying two setups with distinct optical imaging configuration and flow geometry: phase-contrast imaging with Poiseuille flow and confocal microscopy with rheometric flow.

By performing systematic experiments as a function of flow rate and position within the sample, we have investigated the reliability and limits of flow-DDM, established its success over a large range of flow speeds and determined how to optimise imaging parameters. In particular, we have shown that under optimised conditions it is no longer the mean flow speed $\V$ but the width $\Delta v$ of its distribution that limits the reliability of the technique. Therefore, $\Delta v$ should be minimised by imaging away from regions with a large velocity gradient and by ensuring a steady flow.
We have identified an empirical criterion to ensure reliable measurements based on the measured timescales of diffusion and residual velocity, $t_D\lessapprox t_{\Delta v}/3$, which allows estimation of the maximum accessible velocity for reliable measurements, $v_{\max}$ (assuming $\Delta v = k \V$). It is important to note that $v_{\max}$ depends on the particle size; so, based on the measured $t_D$ and $t_{\Delta v}$ obtained from flow-DDM, the above criterion can also be used to give confidence to the user when performing flow-DDM measurements of suspensions with unknown particle-size.

Using the advantages of DDM seen in quiescent systems, flow-DDM allows particle sizing in flowing samples without user inputs or resolution of individual particles (as required for particle tracking), and without the requirement of highly dilute samples (as for DLS). This extends sample possibilities for particle sizing under flow, enabling high-throughput microfluidic testing in development or in-line testing during manufacturing of particulate suspensions, which are so ubiquitous in industry. Moreover, we expect the general framework of flow-DDM to be applicable to other imaging methods, such as bright-field,\cite{trappe2008differential} light-sheet,\cite{wulstein2016light} epifluorescence,\cite{jepson2013enhanced} and dark-field microscopy.\cite{bayles2016dark}

Flow-DDM outperforms current digital Fourier techniques, such as a diffusive fit of anisotropic-DDM\cite{reufer2012differential} or far-field dynamic microscopy.\cite{aime2019probing} Indeed, flow-DDM allows quantitative measurements within $\approx 3\%$ of the free-diffusion coefficient at flow speeds up to one order of magnitude faster than for anisotropic-DDM. Flow-DDM has been designed to be insensitive to the details of the flow, providing some robustness against some spatio-temporal variations. Nevertheless, the method returns measurements of the mean flow velocity and an estimate for the residual velocity spread, which characterises the combination of flow geometry and imaging properties.

Finally, although we have focused entirely on probing diffusive dynamics of dilute suspensions to measure particle size, flow-DDM could also be applied to measure the collective dynamics of dense (and relatively turbid) colloidal suspensions under flow. For example, ready measurement of microscopic particle rearrangements alongside the bulk rheology could bring new insights into the understanding of non-Newtonian fluids such as shear-thickening or yield-stress suspensions\cite{ness2016shear,guy2018constraint} and jammed emulsions.\cite{vasisht2018rate} 

\section*{Conflicts of interest}
There are no conflicts to declare.

\section*{Acknowledgements}
This project has received funding from the European Research Council (ERC) under the European Union's Horizon 2020 research and innovation programme (grant agreement \ftextnumero{s}~731019 and 862559), European Soft Matter Infrastructure (EUSMI) and Novel Characterisation Platform for Formulation Industry (NoChaPFI). The authors thank Andrew Schofield for particle synthesis, and Jean-No\"el Tourvieille and Sophie Galinat for enlightening discussions. All data used are available via Edinburgh DataShare at \url{https://doi.org/10.7488/ds/2987}.

\balance
\providecommand*{\mcitethebibliography}{\thebibliography}
\csname @ifundefined\endcsname{endmcitethebibliography}
{\let\endmcitethebibliography\endthebibliography}{}


\begin{mcitethebibliography}{41}
\providecommand*{\natexlab}[1]{#1}
\providecommand*{\mciteSetBstSublistMode}[1]{}
\providecommand*{\mciteSetBstMaxWidthForm}[2]{}
\providecommand*{\mciteBstWouldAddEndPuncttrue}
  {\def\EndOfBibitem{\unskip.}}
\providecommand*{\mciteBstWouldAddEndPunctfalse}
  {\let\EndOfBibitem\relax}
\providecommand*{\mciteSetBstMidEndSepPunct}[3]{}
\providecommand*{\mciteSetBstSublistLabelBeginEnd}[3]{}
\providecommand*{\EndOfBibitem}{}
\mciteSetBstSublistMode{f}
\mciteSetBstMaxWidthForm{subitem}
{(\emph{\alph{mcitesubitemcount}})}
\mciteSetBstSublistLabelBeginEnd{\mcitemaxwidthsubitemform\space}
{\relax}{\relax}

\bibitem[Bentz \emph{et~al.}(1999)Bentz, Garboczi, Haecker, and
  Jensen]{bentz1999effects}
D.~P. Bentz, E.~J. Garboczi, C.~J. Haecker and O.~M. Jensen,
  \emph{\href{https://doi.org/10.1016/S0008-8846(99)00163-5}{Cem. Concr.
  Res.}}, 1999, \textbf{29}, 1663--1671\relax
\mciteBstWouldAddEndPuncttrue
\mciteSetBstMidEndSepPunct{\mcitedefaultmidpunct}
{\mcitedefaultendpunct}{\mcitedefaultseppunct}\relax
\EndOfBibitem
\bibitem[Kan(1999)]{kan1999role}
C.~S. Kan, \emph{\href{https://doi.org/10.1007/BF02698388}{J. Coat. Technol.
  Res.}}, 1999, \textbf{71}, 89\relax
\mciteBstWouldAddEndPuncttrue
\mciteSetBstMidEndSepPunct{\mcitedefaultmidpunct}
{\mcitedefaultendpunct}{\mcitedefaultseppunct}\relax
\EndOfBibitem
\bibitem[Sandri \emph{et~al.}(2014)Sandri, Bonferoni, Ferrari, Rossi, and
  Caramella]{sandri2014role}
G.~Sandri, M.~C. Bonferoni, F.~Ferrari, S.~Rossi and C.~M. Caramella,
  \emph{\href{https://doi.org/10.1007/978-3-319-00714-4_11}{Particulate
  Products}}, Springer, 2014, pp. 323--341\relax
\mciteBstWouldAddEndPuncttrue
\mciteSetBstMidEndSepPunct{\mcitedefaultmidpunct}
{\mcitedefaultendpunct}{\mcitedefaultseppunct}\relax
\EndOfBibitem
\bibitem[Ziegler \emph{et~al.}(2001)Ziegler, Mongia, and
  Hollender]{ziegler2001role}
G.~R. Ziegler, G.~Mongia and R.~Hollender,
  \emph{\href{https://doi.org/10.1080/10942910009524621}{Int. J. Food Prop.}},
  2001, \textbf{4}, 353--370\relax
\mciteBstWouldAddEndPuncttrue
\mciteSetBstMidEndSepPunct{\mcitedefaultmidpunct}
{\mcitedefaultendpunct}{\mcitedefaultseppunct}\relax
\EndOfBibitem
\bibitem[Bell \emph{et~al.}(2012)Bell, Minelli, Tompkins, Stevens, and
  Shard]{bell2012emerging}
N.~C. Bell, C.~Minelli, J.~Tompkins, M.~M. Stevens and A.~G. Shard,
  \emph{\href{https://doi.org/10.1021/la301351k}{Langmuir}}, 2012, \textbf{28},
  10860--10872\relax
\mciteBstWouldAddEndPuncttrue
\mciteSetBstMidEndSepPunct{\mcitedefaultmidpunct}
{\mcitedefaultendpunct}{\mcitedefaultseppunct}\relax
\EndOfBibitem
\bibitem[Berne and Pecora(2000)]{berne2000dynamic}
B.~J. Berne and R.~Pecora, \emph{Dynamic light scattering: with applications to
  chemistry, biology, and physics}, Dover Publications, Mineola (N.Y.),
  2000\relax
\mciteBstWouldAddEndPuncttrue
\mciteSetBstMidEndSepPunct{\mcitedefaultmidpunct}
{\mcitedefaultendpunct}{\mcitedefaultseppunct}\relax
\EndOfBibitem
\bibitem[Leung \emph{et~al.}(2006)Leung, Suh, and Ansari]{leung2006particle}
A.~B. Leung, K.~I. Suh and R.~R. Ansari,
  \emph{\href{https://doi.org/10.1364/AO.45.002186}{Appl. Opt.}}, 2006,
  \textbf{45}, 2186--2190\relax
\mciteBstWouldAddEndPuncttrue
\mciteSetBstMidEndSepPunct{\mcitedefaultmidpunct}
{\mcitedefaultendpunct}{\mcitedefaultseppunct}\relax
\EndOfBibitem
\bibitem[Pusey(1999)]{pusey1999suppression}
P.~Pusey,
  \emph{\href{http://www.sciencedirect.com/science/article/pii/S1359029499000369}{Curr.
  Opin. Colloid Interface Sci.}}, 1999, \textbf{4}, 177 -- 185\relax
\mciteBstWouldAddEndPuncttrue
\mciteSetBstMidEndSepPunct{\mcitedefaultmidpunct}
{\mcitedefaultendpunct}{\mcitedefaultseppunct}\relax
\EndOfBibitem
\bibitem[Urban and Schurtenberger(1998)]{urban1998characterization}
C.~Urban and P.~Schurtenberger,
  \emph{\href{https://doi.org/10.1006/jcis.1998.5769}{J. Colloid Interface
  Sci.}}, 1998, \textbf{207}, 150 -- 158\relax
\mciteBstWouldAddEndPuncttrue
\mciteSetBstMidEndSepPunct{\mcitedefaultmidpunct}
{\mcitedefaultendpunct}{\mcitedefaultseppunct}\relax
\EndOfBibitem
\bibitem[Finder \emph{et~al.}(2004)Finder, Wohlgemuth, and
  Mayer]{finder2004analysis}
C.~Finder, M.~Wohlgemuth and C.~Mayer,
  \emph{\href{https://doi.org/10.1002/ppsc.200400948}{Part. Part. Syst.
  Charact.}}, 2004, \textbf{21}, 372--378\relax
\mciteBstWouldAddEndPuncttrue
\mciteSetBstMidEndSepPunct{\mcitedefaultmidpunct}
{\mcitedefaultendpunct}{\mcitedefaultseppunct}\relax
\EndOfBibitem
\bibitem[Crocker and Grier(1996)]{crocker1996methods}
J.~C. Crocker and D.~G. Grier,
  \emph{\href{https://doi.org/10.1006/jcis.1996.0217}{J. Colloid Interface
  Sci.}}, 1996, \textbf{179}, 298--310\relax
\mciteBstWouldAddEndPuncttrue
\mciteSetBstMidEndSepPunct{\mcitedefaultmidpunct}
{\mcitedefaultendpunct}{\mcitedefaultseppunct}\relax
\EndOfBibitem
\bibitem[Newby \emph{et~al.}(2018)Newby, Schaefer, Lee, Forest, and
  Lai]{newby2018convolutional}
J.~M. Newby, A.~M. Schaefer, P.~T. Lee, M.~G. Forest and S.~K. Lai,
  \emph{\href{https://www.pnas.org/content/115/36/9026}{Proc. Natl. Acad. Sci.
  U.S.A.}}, 2018, \textbf{115}, 9026--9031\relax
\mciteBstWouldAddEndPuncttrue
\mciteSetBstMidEndSepPunct{\mcitedefaultmidpunct}
{\mcitedefaultendpunct}{\mcitedefaultseppunct}\relax
\EndOfBibitem
\bibitem[Cerbino and Trappe(2008)]{trappe2008differential}
R.~Cerbino and V.~Trappe,
  \emph{\href{https://link.aps.org/doi/10.1103/PhysRevLett.100.188102}{Phys.
  Rev. Lett.}}, 2008, \textbf{100}, 188102\relax
\mciteBstWouldAddEndPuncttrue
\mciteSetBstMidEndSepPunct{\mcitedefaultmidpunct}
{\mcitedefaultendpunct}{\mcitedefaultseppunct}\relax
\EndOfBibitem
\bibitem[Edera \emph{et~al.}(2017)Edera, Bergamini, Trappe, Giavazzi, and
  Cerbino]{edera2017differential}
P.~Edera, D.~Bergamini, V.~Trappe, F.~Giavazzi and R.~Cerbino,
  \emph{\href{https://link.aps.org/doi/10.1103/PhysRevMaterials.1.073804}{Phys.
  Rev. Materials}}, 2017, \textbf{1}, 073804\relax
\mciteBstWouldAddEndPuncttrue
\mciteSetBstMidEndSepPunct{\mcitedefaultmidpunct}
{\mcitedefaultendpunct}{\mcitedefaultseppunct}\relax
\EndOfBibitem
\bibitem[Bayles \emph{et~al.}(2017)Bayles, Squires, and
  Helgeson]{bayles2017probe}
A.~V. Bayles, T.~M. Squires and M.~E. Helgeson,
  \emph{\href{https://doi.org/10.1007/s00397-017-1047-7}{Rheol. Acta}}, 2017,
  \textbf{56}, 863--869\relax
\mciteBstWouldAddEndPuncttrue
\mciteSetBstMidEndSepPunct{\mcitedefaultmidpunct}
{\mcitedefaultendpunct}{\mcitedefaultseppunct}\relax
\EndOfBibitem
\bibitem[Escobedo-Sánchez \emph{et~al.}(2018)Escobedo-Sánchez,
  Segovia-Gutiérrez, Zuccolotto-Bernez, Hansen, Marciniak, Sachowsky, Platten,
  and Egelhaaf]{sanchez2018microliter}
M.~A. Escobedo-Sánchez, J.~P. Segovia-Gutiérrez, A.~B. Zuccolotto-Bernez,
  J.~Hansen, C.~C. Marciniak, K.~Sachowsky, F.~Platten and S.~U. Egelhaaf,
  \emph{\href{http://dx.doi.org/10.1039/C8SM00784E}{Soft Matter}}, 2018,
  \textbf{14}, 7016--7025\relax
\mciteBstWouldAddEndPuncttrue
\mciteSetBstMidEndSepPunct{\mcitedefaultmidpunct}
{\mcitedefaultendpunct}{\mcitedefaultseppunct}\relax
\EndOfBibitem
\bibitem[Wilson \emph{et~al.}(2011)Wilson, Martinez, Schwarz-Linek, Tailleur,
  Bryant, Pusey, and Poon]{wilson2011differential}
L.~G. Wilson, V.~A. Martinez, J.~Schwarz-Linek, J.~Tailleur, G.~Bryant, P.~N.
  Pusey and W.~C.~K. Poon,
  \emph{\href{https://link.aps.org/doi/10.1103/PhysRevLett.106.018101}{Phys.
  Rev. Lett.}}, 2011, \textbf{106}, 018101\relax
\mciteBstWouldAddEndPuncttrue
\mciteSetBstMidEndSepPunct{\mcitedefaultmidpunct}
{\mcitedefaultendpunct}{\mcitedefaultseppunct}\relax
\EndOfBibitem
\bibitem[Martinez \emph{et~al.}(2012)Martinez, Besseling, Croze, Tailleur,
  Reufer, Schwarz-Linek, Wilson, Bees, and Poon]{martinez2012differential}
V.~A. Martinez, R.~Besseling, O.~A. Croze, J.~Tailleur, M.~Reufer,
  J.~Schwarz-Linek, L.~G. Wilson, M.~A. Bees and W.~C.~K. Poon,
  \emph{\href{http://www.sciencedirect.com/science/article/pii/S0006349512009721}{Biophys.
  J}}, 2012, \textbf{103}, 1637 -- 1647\relax
\mciteBstWouldAddEndPuncttrue
\mciteSetBstMidEndSepPunct{\mcitedefaultmidpunct}
{\mcitedefaultendpunct}{\mcitedefaultseppunct}\relax
\EndOfBibitem
\bibitem[Jepson \emph{et~al.}(2019)Jepson, Arlt, Statham, Spilman, Burton,
  Wood, Poon, and Martinez]{jepson2019high}
A.~Jepson, J.~Arlt, J.~Statham, M.~Spilman, K.~Burton, T.~Wood, W.~C.~K. Poon
  and V.~A. Martinez,
  \emph{\href{https://doi.org/10.1371/journal.pone.0202720}{PLOS ONE}}, 2019,
  \textbf{14}, 1--17\relax
\mciteBstWouldAddEndPuncttrue
\mciteSetBstMidEndSepPunct{\mcitedefaultmidpunct}
{\mcitedefaultendpunct}{\mcitedefaultseppunct}\relax
\EndOfBibitem
\bibitem[Latreille \emph{et~al.}(2019)Latreille, Adibnia, Nour, Rabanel,
  Lalloz, Arlt, Poon, Hildgen, Martinez, and Banquy]{latreille2019spontaneous}
P.-L. Latreille, V.~Adibnia, A.~Nour, J.-M. Rabanel, A.~Lalloz, J.~Arlt,
  W.~C.~K. Poon, P.~Hildgen, V.~A. Martinez and X.~Banquy,
  \emph{\href{https://doi.org/10.1038/s41467-019-12246-x}{Nat. Commun.}}, 2019,
  \textbf{10}, 1--8\relax
\mciteBstWouldAddEndPuncttrue
\mciteSetBstMidEndSepPunct{\mcitedefaultmidpunct}
{\mcitedefaultendpunct}{\mcitedefaultseppunct}\relax
\EndOfBibitem
\bibitem[Regan \emph{et~al.}(2019)Regan, Wulstein, Rasmussen, McGorty, and
  Robertson-Anderson]{regan2019bridging}
K.~Regan, D.~Wulstein, H.~Rasmussen, R.~McGorty and R.~M. Robertson-Anderson,
  \emph{\href{https://doi.org/10.1039/C8SM02023J}{Soft Matter}}, 2019,
  \textbf{15}, 1200--1209\relax
\mciteBstWouldAddEndPuncttrue
\mciteSetBstMidEndSepPunct{\mcitedefaultmidpunct}
{\mcitedefaultendpunct}{\mcitedefaultseppunct}\relax
\EndOfBibitem
\bibitem[Reufer \emph{et~al.}(2012)Reufer, Martinez, Schurtenberger, and
  Poon]{reufer2012differential}
M.~Reufer, V.~A. Martinez, P.~Schurtenberger and W.~C.~K. Poon,
  \emph{\href{https://doi.org/10.1021/la204904a}{Langmuir}}, 2012, \textbf{28},
  4618--4624\relax
\mciteBstWouldAddEndPuncttrue
\mciteSetBstMidEndSepPunct{\mcitedefaultmidpunct}
{\mcitedefaultendpunct}{\mcitedefaultseppunct}\relax
\EndOfBibitem
\bibitem[Lu \emph{et~al.}(2012)Lu, Giavazzi, Angelini, Zaccarelli, Jargstorff,
  Schofield, Wilking, Romanowsky, Weitz, and Cerbino]{lu2012characterizing}
P.~J. Lu, F.~Giavazzi, T.~E. Angelini, E.~Zaccarelli, F.~Jargstorff, A.~B.
  Schofield, J.~N. Wilking, M.~B. Romanowsky, D.~A. Weitz and R.~Cerbino,
  \emph{\href{https://link.aps.org/doi/10.1103/PhysRevLett.108.218103}{Phys.
  Rev. Lett.}}, 2012, \textbf{108}, 218103\relax
\mciteBstWouldAddEndPuncttrue
\mciteSetBstMidEndSepPunct{\mcitedefaultmidpunct}
{\mcitedefaultendpunct}{\mcitedefaultseppunct}\relax
\EndOfBibitem
\bibitem[L\'azaro-L\'azaro \emph{et~al.}(2019)L\'azaro-L\'azaro, Perera-Burgos,
  Laermann, Sentjabrskaja, P\'erez-\'Angel, Laurati, Egelhaaf, Medina-Noyola,
  Voigtmann, Casta\~neda Priego, and Elizondo-Aguilera]{lazaro2019glassy}
E.~L\'azaro-L\'azaro, J.~A. Perera-Burgos, P.~Laermann, T.~Sentjabrskaja,
  G.~P\'erez-\'Angel, M.~Laurati, S.~U. Egelhaaf, M.~Medina-Noyola,
  T.~Voigtmann, R.~Casta\~neda Priego and L.~F. Elizondo-Aguilera,
  \emph{\href{https://link.aps.org/doi/10.1103/PhysRevE.99.042603}{Phys. Rev.
  E}}, 2019, \textbf{99}, 042603\relax
\mciteBstWouldAddEndPuncttrue
\mciteSetBstMidEndSepPunct{\mcitedefaultmidpunct}
{\mcitedefaultendpunct}{\mcitedefaultseppunct}\relax
\EndOfBibitem
\bibitem[Pal \emph{et~al.}(2020)Pal, Martinez, Ito, Arlt, Crassous, Poon, and
  Schurtenberger]{pal2020anisotropic}
A.~Pal, V.~A. Martinez, T.~H. Ito, J.~Arlt, J.~J. Crassous, W.~C.~K. Poon and
  P.~Schurtenberger,
  \emph{\href{https://advances.sciencemag.org/content/6/3/eaaw9733}{Sci.
  Adv.}}, 2020, \textbf{6}, eaaw9733\relax
\mciteBstWouldAddEndPuncttrue
\mciteSetBstMidEndSepPunct{\mcitedefaultmidpunct}
{\mcitedefaultendpunct}{\mcitedefaultseppunct}\relax
\EndOfBibitem
\bibitem[Tong \emph{et~al.}(2016)Tong, Brown, Stone, Cree, and
  Chamley]{tong2016flow}
M.~Tong, O.~S. Brown, P.~R. Stone, L.~M. Cree and L.~W. Chamley,
  \emph{\href{https://doi.org/10.1016/j.placenta.2015.12.004}{Placenta}}, 2016,
  \textbf{38}, 29--32\relax
\mciteBstWouldAddEndPuncttrue
\mciteSetBstMidEndSepPunct{\mcitedefaultmidpunct}
{\mcitedefaultendpunct}{\mcitedefaultseppunct}\relax
\EndOfBibitem
\bibitem[Philippe \emph{et~al.}(2016)Philippe, Aime, Roger, Jelinek,
  Pr{\'{e}}vot, Berthier, and Cipelletti]{philippe2016efficient}
A.~Philippe, S.~Aime, V.~Roger, R.~Jelinek, G.~Pr{\'{e}}vot, L.~Berthier and
  L.~Cipelletti, \emph{\href{https://doi.org/10.1088/0953-8984/28/7/075201}{J.
  Phys. Condens. Matter}}, 2016, \textbf{28}, 075201\relax
\mciteBstWouldAddEndPuncttrue
\mciteSetBstMidEndSepPunct{\mcitedefaultmidpunct}
{\mcitedefaultendpunct}{\mcitedefaultseppunct}\relax
\EndOfBibitem
\bibitem[Aime and Cipelletti(2019)]{aime2019probing}
S.~Aime and L.~Cipelletti,
  \emph{\href{http://dx.doi.org/10.1039/C8SM01564C}{Soft Matter}}, 2019,
  \textbf{15}, 213--226\relax
\mciteBstWouldAddEndPuncttrue
\mciteSetBstMidEndSepPunct{\mcitedefaultmidpunct}
{\mcitedefaultendpunct}{\mcitedefaultseppunct}\relax
\EndOfBibitem
\bibitem[Besseling \emph{et~al.}(2009)Besseling, Isa, Weeks, and
  Poon]{besseling2009flow}
R.~Besseling, L.~Isa, E.~R. Weeks and W.~C.~K. Poon,
  \emph{\href{https://doi.org/10.1016/j.cis.2008.09.008}{Adv. Colloid Interface
  Sci.}}, 2009, \textbf{146}, 1--17\relax
\mciteBstWouldAddEndPuncttrue
\mciteSetBstMidEndSepPunct{\mcitedefaultmidpunct}
{\mcitedefaultendpunct}{\mcitedefaultseppunct}\relax
\EndOfBibitem
\bibitem[Colin \emph{et~al.}(2014)Colin, Zhang, and Wilson]{colin2014fast}
R.~Colin, R.~Zhang and L.~G. Wilson,
  \emph{\href{https://doi.org/10.1098/rsif.2014.0486}{J. R. Soc. Interface}},
  2014, \textbf{11}, 20140486\relax
\mciteBstWouldAddEndPuncttrue
\mciteSetBstMidEndSepPunct{\mcitedefaultmidpunct}
{\mcitedefaultendpunct}{\mcitedefaultseppunct}\relax
\EndOfBibitem
\bibitem[Giavazzi \emph{et~al.}(2009)Giavazzi, Brogioli, Trappe, Bellini, and
  Cerbino]{giavazzi2009scattering}
F.~Giavazzi, D.~Brogioli, V.~Trappe, T.~Bellini and R.~Cerbino,
  \emph{\href{https://link.aps.org/doi/10.1103/PhysRevE.80.031403}{Phys. Rev.
  E}}, 2009, \textbf{80}, 031403\relax
\mciteBstWouldAddEndPuncttrue
\mciteSetBstMidEndSepPunct{\mcitedefaultmidpunct}
{\mcitedefaultendpunct}{\mcitedefaultseppunct}\relax
\EndOfBibitem
\bibitem[Besseling \emph{et~al.}(2007)Besseling, Weeks, Schofield, and
  Poon]{besseling2007three}
R.~Besseling, E.~R. Weeks, A.~B. Schofield and W.~C.~K. Poon,
  \emph{\href{https://link.aps.org/doi/10.1103/PhysRevLett.99.028301}{Phys.
  Rev. Lett.}}, 2007, \textbf{99}, 028301\relax
\mciteBstWouldAddEndPuncttrue
\mciteSetBstMidEndSepPunct{\mcitedefaultmidpunct}
{\mcitedefaultendpunct}{\mcitedefaultseppunct}\relax
\EndOfBibitem
\bibitem[Giavazzi \emph{et~al.}(2017)Giavazzi, Edera, Lu, and
  Cerbino]{giavazzi2017image}
F.~Giavazzi, P.~Edera, P.~J. Lu and R.~Cerbino,
  \emph{\href{https://doi.org/10.1140/epje/i2017-11587-3}{Eur. Phys. J. E}},
  2017, \textbf{40}, 97\relax
\mciteBstWouldAddEndPuncttrue
\mciteSetBstMidEndSepPunct{\mcitedefaultmidpunct}
{\mcitedefaultendpunct}{\mcitedefaultseppunct}\relax
\EndOfBibitem
\bibitem[Boussinesq(1868)]{boussinesq1868memoire}
J.~Boussinesq, \emph{J. Math Pures. Appl.}, 1868,  377--424\relax
\mciteBstWouldAddEndPuncttrue
\mciteSetBstMidEndSepPunct{\mcitedefaultmidpunct}
{\mcitedefaultendpunct}{\mcitedefaultseppunct}\relax
\EndOfBibitem
\bibitem[Dudgeon and Wedgewood(1993)]{dudgeon1993flow}
D.~J. Dudgeon and L.~E. Wedgewood,
  \emph{\href{http://dx.doi.org/10.1016/0377-0257(93)80063-H}{J Nonnewton.
  Fluid Mech.}}, 1993, \textbf{48}, 21 -- 48\relax
\mciteBstWouldAddEndPuncttrue
\mciteSetBstMidEndSepPunct{\mcitedefaultmidpunct}
{\mcitedefaultendpunct}{\mcitedefaultseppunct}\relax
\EndOfBibitem
\bibitem[Wulstein \emph{et~al.}(2016)Wulstein, Regan, Robertson-Anderson, and
  McGorty]{wulstein2016light}
D.~M. Wulstein, K.~E. Regan, R.~M. Robertson-Anderson and R.~McGorty,
  \emph{\href{http://dx.doi.org/10.1364/OE.24.020881}{Optics Express}}, 2016,
  \textbf{24}, 20881--20894\relax
\mciteBstWouldAddEndPuncttrue
\mciteSetBstMidEndSepPunct{\mcitedefaultmidpunct}
{\mcitedefaultendpunct}{\mcitedefaultseppunct}\relax
\EndOfBibitem
\bibitem[Jepson \emph{et~al.}(2013)Jepson, Martinez, Schwarz-Linek, Morozov,
  and Poon]{jepson2013enhanced}
A.~Jepson, V.~A. Martinez, J.~Schwarz-Linek, A.~Morozov and W.~C.~K. Poon,
  \emph{\href{https://doi.org/10.1103/PhysRevE.88.041002}{Phys. Rev. E}}, 2013,
  \textbf{88}, 041002\relax
\mciteBstWouldAddEndPuncttrue
\mciteSetBstMidEndSepPunct{\mcitedefaultmidpunct}
{\mcitedefaultendpunct}{\mcitedefaultseppunct}\relax
\EndOfBibitem
\bibitem[Bayles \emph{et~al.}(2016)Bayles, Squires, and
  Helgeson]{bayles2016dark}
A.~V. Bayles, T.~M. Squires and M.~E. Helgeson,
  \emph{\href{http://dx.doi.org/10.1039/C5SM02576A}{Soft Matter}}, 2016,
  \textbf{12}, 2440--2452\relax
\mciteBstWouldAddEndPuncttrue
\mciteSetBstMidEndSepPunct{\mcitedefaultmidpunct}
{\mcitedefaultendpunct}{\mcitedefaultseppunct}\relax
\EndOfBibitem
\bibitem[Ness and Sun(2016)]{ness2016shear}
C.~Ness and J.~Sun, \emph{\href{http://dx.doi.org/10.1039/C5SM02326B}{Soft
  Matter}}, 2016, \textbf{12}, 914--924\relax
\mciteBstWouldAddEndPuncttrue
\mciteSetBstMidEndSepPunct{\mcitedefaultmidpunct}
{\mcitedefaultendpunct}{\mcitedefaultseppunct}\relax
\EndOfBibitem
\bibitem[Guy \emph{et~al.}(2018)Guy, Richards, Hodgson, Blanco, and
  Poon]{guy2018constraint}
B.~M. Guy, J.~A. Richards, D.~J.~M. Hodgson, E.~Blanco and W.~C.~K. Poon,
  \emph{\href{https://link.aps.org/doi/10.1103/PhysRevLett.121.128001}{Phys.
  Rev. Lett.}}, 2018, \textbf{121}, 128001\relax
\mciteBstWouldAddEndPuncttrue
\mciteSetBstMidEndSepPunct{\mcitedefaultmidpunct}
{\mcitedefaultendpunct}{\mcitedefaultseppunct}\relax
\EndOfBibitem
\bibitem[Vasisht \emph{et~al.}(2018)Vasisht, Dutta, Del~Gado, and
  Blair]{vasisht2018rate}
V.~V. Vasisht, S.~K. Dutta, E.~Del~Gado and D.~L. Blair,
  \emph{\href{https://link.aps.org/doi/10.1103/PhysRevLett.120.018001}{Phys.
  Rev. Lett.}}, 2018, \textbf{120}, 018001\relax
\mciteBstWouldAddEndPuncttrue
\mciteSetBstMidEndSepPunct{\mcitedefaultmidpunct}
{\mcitedefaultendpunct}{\mcitedefaultseppunct}\relax
\EndOfBibitem
\end{mcitethebibliography}
\end{document}